\documentclass{aa}  

\usepackage{graphicx}
\usepackage{txfonts}
\usepackage{epsfig}
\usepackage{longtable}
\usepackage{amssymb} 
\usepackage[colorlinks,allcolors=blue]{hyperref}
\begin{document}

   \title{Properties and observability 
          of glitches and anti-glitches\\ in accreting pulsars}
   \titlerunning{Glitches and anti-glitches in accreting pulsars}
   
   \author{L. Ducci
          \inst{1,2}
          \and 
          P. M. Pizzochero\inst{3,4}
          \and
          V. Doroshenko\inst{1}
          \and
          A. Santangelo\inst{1}
          \and
          S. Mereghetti\inst{5}
          \and
          C. Ferrigno\inst{2}
          }

   \institute{Institut f\"ur Astronomie und Astrophysik, Eberhard Karls Universit\"at, 
              Sand 1, 72076 T\"ubingen, Germany\\
              \email{ducci@astro.uni-tuebingen.de}
              \and
              ISDC Data Center for Astrophysics, Universit\'e de Gen\`eve, 16 chemin d'\'Ecogia, 1290 Versoix, Switzerland
              \and
              Dipartimento di Fisica, Universit\`a degli Studi di Milano, Via Celoria 16, 20133 Milano, Italy
              \and
              Istituto Nazionale di Fisica Nucleare, sezione di Milano, Via Celoria 16, 20133 Milano, Italy
              \and
              INAF -- Istituto di Astrofisica Spaziale e Fisica Cosmica, Via E. Bassini 15, 20133 Milano, Italy
             }

   \date{Received ...; accepted ...}

 
  \abstract
   {Several glitches have been observed in young, isolated radio pulsars, while a clear detection in
    accretion-powered X-ray pulsars is still lacking.
    We use the Pizzochero snowplow model for pulsar glitches as well as starquake models
    to determine for the first time the expected properties of glitches 
    in accreting pulsars and their observability. Since some accreting pulsars 
    show accretion-induced long-term spin-up, we also investigate the possibility 
    that anti-glitches occur in these stars.
    We find that glitches caused by quakes in a slow accreting neutron star
    are very rare and their detection extremely unlikely. 
    On the contrary, glitches and anti-glitches caused by a transfer of angular momentum 
    between the superfluid neutron vortices and the non-superfluid component
    may take place in accreting pulsars more often.
    We calculate the maximum jump
    in angular velocity of an anti-glitch and we find that it is
    expected to be $\Delta \Omega_{\rm a-gl}\approx 10^{-5}-10^{-4}$~rad~s$^{-1}$.
    We also note that since accreting pulsars usually have rotational angular velocities
    lower than those of isolated glitching pulsars, 
    both glitches and anti-glitches are expected to have long rise 
    and recovery  timescales compared to isolated glitching pulsars,
    with glitches and anti-glitches appearing as a simple step in angular velocity.
    Among accreting pulsars,  we find that  GX~1+4 is the best candidate for the detection of glitches
    with currently operating X-ray instruments and future missions such as the proposed
    \emph{Large Observatory for X-ray Timing (LOFT)}.}


   \maketitle
%

\begin{table*}
\begin{center}
\caption{Rotational parameters and luminosities of four accreting pulsars showing among the largest
and stable spin-down rates ($\dot{\Omega}_\infty$) in XRBs. $\Omega$ is the rotational angular velocity,
$\tau_\infty$ is the torque reversal timescale, and $L_x$ is the X-ray luminosity.}
\label{Table}
\begin{tabular}{lcccccccc}
\hline
\hline          
\noalign{\smallskip}
      Name     &\multicolumn{2}{c}{$\Omega$}& \multicolumn{2}{c}{$\dot{\Omega}_\infty$} & \multicolumn{2}{c}{$\tau_\infty$} & \multicolumn{2}{c}{$L_x$}       \\
               &\multicolumn{2}{c}{rad~s$^{-1}$}&  \multicolumn{2}{c}{rad~s$^{-2}$}    &     \multicolumn{2}{c}{d}        & \multicolumn{2}{c}{erg~s$^{-1}$ } \\ 
\noalign{\smallskip}
\hline 
\noalign{\smallskip}                                                                    
GX~1+4         & $\approx 0.042$      & (1)  & $-3 \times 10^{-11}$         &  (1)  &          $>9900$          & (1) & $10^{35}-10^{36}$ & (1)  \\
\noalign{\smallskip}
OAO~1657$-$415 & $\approx 0.17$        & (2)  & $-1.3 \times 10^{-11}$       &  (2)  &        $\approx 200$      & (3) & $10^{36}-10^{37}$ & (2)  \\
\noalign{\smallskip}
4U~1626$-$67   & $\approx 0.8$        & (4)  & $-4.4 \times 10^{-12}$       &  (4)  &        $\approx 6600$     & (5) & $\gtrsim 10^{36}$ & (4) \\
\noalign{\smallskip}
4U~1907+09     & $\approx 0.014$      & (6)  & $-2.3 \times 10^{-13}$       &  (6)  &          $>5500$          & (6) &     $10^{36}$     & (7)  \\
\noalign{\smallskip}
\hline
\end{tabular}
\end{center}
References: \\
(1) \citet{Gonzalez-Galan12}; (2) \citet{Jenke12}; (3) \citet{Barnstedt08}; (4) \citet{Chakrabarty97}; (5) \citet{Beri14};
(6) \citet{Inam09}; (7) \citet{intZand97}.
\end{table*}

\section{Introduction}
\label{sect. intro}

Isolated pulsars are considered remarkably stable rotators showing
long-term spin-down caused by the emission of electromagnetic dipole
radiation, relativistic outflows, and possibly gravitational waves \citep{LIGO14}.
However,  they have sometimes  been observed to glitch.
Glitches are sudden increases in the spin rate followed
by a relaxation toward a steadily, long-term, decreasing spin
rate.
They have been observed in over 100 isolated radio pulsars and magnetars
(see, e.g., \citealt{Espinoza11}; \citealt{Yu13}; \citealt{Dib08}),
with jumps in angular velocity up to $\Delta \Omega_{\rm gl} \approx 10^{-4}$~rad~s$^{-1}$.

There are two main models to explain glitches:
the \emph{starquake glitch model} (e.g., \citealt{Baym71})
and the \emph{superfluid vortex unpinning model} 
(\citealt{Anderson75}; see also \citealt{Haskell15} for a recent review
on models of pulsar glitches).
In the starquake glitch model of \citet{Baym71}, 
a rapidly rotating pulsar has an oblate crust that deforms toward an almost
spherical shape as the pulsar slows down.
This leads to a sudden crack in the crust and a decrease in the 
moment of inertia resulting in a sudden increase 
in the angular velocity of the star.
In the model proposed by \citet{Anderson75} the glitch is caused
by a sudden unpinning of neutron superfluid vortices from the lattice nuclei
in different regions of the interior of a neutron star (NS).
The angular momentum stored in the superfluid vortices is then transferred 
to the non-superfluid component of the star, leading to the observed 
fast spin-up of the NS rotation angular velocity.

No clear observational evidence that accreting NSs 
experience glitches has been presented so far.
Nonetheless, they have been proposed as a possible explanation 
of the peculiar timing behavior of some accreting pulsars.
\citet{Galloway04} observed a sudden spin-up (with a fractional change in frequency of
$\sim 3.7\times10^{-5}$ in $\lesssim 10$~hr) in the accreting 18.7~s pulsar KS~1947+300
with \emph{RXTE}, that was interpreted as a glitch event.
Contrary to what happens in the classical
glitch scenario, the sudden spin-up was observed when the NS
was already spinning up because of the transfer of angular momentum
from the accretion disk.
\citet{Galloway04} did not rule out the possibility that the 
sudden spin-up was caused by a short episode
of enhanced accretion.
\citet{Klochkov09} studied the pulse period variations of Her~X$-$1 using
\emph{Swift}/BAT data. They found two spin-down episodes followed by spin-up
intervals where the spin period increased exponentially,
a phenomenon typical of the post-glitch recovery stage, but with an opposite sign.

Observations of glitching pulsars showed that the glitch frequency increases
with the long-term spin-down rate \citep{Espinoza11}.
Since the spin-down caused by electromagnetic braking
decreases as pulsars age, the detection of glitches in old
pulsars ($>10^4$~yr, \citealt{Espinoza11}), such as accreting pulsars in X-ray binaries,
is expected to drop.
In this paper we  show that glitch frequency in accreting pulsars
in X-ray binaries can  actually be higher than expected for old pulsars.
The aim of this paper is to assess the conditions under which glitches are more likely
to occur in an accreting pulsar and investigate, for the first time, 
the expected properties of glitches in these objects.

In Sect. \ref{Sect. calculations} we calculate the expected jump in angular velocity
and the interglitch interval in the starquake and superfluid vortex scenarios,
assuming that glitches occur during a long-term spin-down.
Then, we investigate the properties of anti-glitches (sudden spin-down)
during long-term spin-up of the NS. We modified the model of \citet{Pizzochero11}
to calculate the jump in angular velocity in this scenario.
In Sect. \ref{sect. observability} we discuss the observability of glitches
and anti-glitches in accreting X-ray pulsars.

\section{Properties of glitches in accreting pulsars}
\label{Sect. calculations}         

Pulsars in X-ray binary systems (XRBs) 
can experience spin-up and spin-down caused by
the interaction between the accretion flow and
the magnetosphere of the NS (\citealt{Pringle72}; \citealt{Rappaport77}).
The interaction may occur directly from the wind of the donor star or through
an accretion disk (e.g., \citealt{Bildsten97} and references therein).
In some XRBs (see  examples in Table \ref{Table})
the accretion torque is responsible for variations
of the angular velocity of the NS of magnitude comparable to that caused
by the electromagnetic braking torque in young, glitching pulsars
($10^{-15}-10^{-10}$~rad~s$^{-2}$, \citealt{Lyne00} and, for comparison, Table \ref{Table}).
Since the rate of glitches increases with the 
spin-down rate (see following sections), glitches in XRBs
should be quite frequent. In the next  sections we derive 
their main properties in different scenarios.

\subsection{Starquake glitch scenario}

The Baym-Pines model allows  the time $\Delta t_{\rm q}$ 
to the next quake to be estimated from the magnitude of the preceeding one through the equation
\begin{equation} \label{deltat baym}
\Delta t_{\rm q} = T \frac{2A^2}{B I_0 \Omega^2} | \Delta \epsilon | \mbox{ ,}
\end{equation}
where $T \equiv - \Omega / \dot{\Omega}_\infty$ 
is the characteristic timescale at which 
the pulsar slows down because of the loss of rotational energy,
$\Omega$ is the rotational angular velocity of the NS,
$\dot{\Omega}_\infty$ is the long-term spin-down rate,
$| \Delta \epsilon |$ is the reduction in oblateness $\epsilon$ caused
by the quake in the previous glitch, $A$ and $B$ are coefficients
describing respectively the gravitational and elastic energy 
stored in the NS as a result of its rotation,
and $I_0$ is the moment of inertia. 
The values of $A$, $B$, and $I_0$ depend on the equation of state (EoS)
of the neutron star core and on the crust model adopted.
Here we adopt the parameters obtained by \citet{Pandharipande76}
and \citet{Zdunik08}, which updated the Baym-Pines model
using different EoSs and crust models.

We assume different values of $\Delta \epsilon$
ranging from $10^{-10}$ to $10^{-6}$ (which are the typical values
of glitches observed in isolated pulsars).
We use $\Omega$ and $\dot{\Omega}_\infty$ of the four XRBs in 
Table \ref{Table}. 
We find that for any value of $\Omega$, $\dot{\Omega}_\infty$, $A$, $B$, 
$I_0$, and $\Delta \epsilon$, 
$\Delta t_{\rm q}$ is always greater than $10^5$~yr. The value of
$\Delta t_{\rm q}$ can be $\ll 10^5$~yr for $\Delta \epsilon \ll 10^{-10}$.
However, starquakes with such small $\Delta \epsilon$ would be difficult to detect.
Therefore, the detection of a glitch produced by a starquake in an accreting pulsar
is extremely unlikely.
Moreover, \citet{Zdunik08} showed that an accreted crust is softer 
to an elastic deformation than the crust of an isolated 
pulsar\footnote{\citet{Zdunik08} pointed out that the calculations 
for the elastic properties of accreted crusts
are subject to high uncertainties, and therefore they
must be taken with caution.}.
As a consequence, $\Delta t_{\rm q}$ would be longer and
glitches caused by quakes  rarer.

\subsection{Superfluid vortex scenario}
\label{sect case2}

Our calculation of the interval between glitches $\Delta t_{\rm gl}$
and of the jump in angular velocity $\Delta \Omega_{\rm gl}$ of accreting pulsars
in the superfluid vortex scenario is based on the ``snowplow'' model
of \citet{Pizzochero11}.
In this model, the matter of the NS is divided into two components:
the neutron superfluid and the normal component (charged components that corotate
with the pulsar magnetic field). 
In the core ($r<R_{\rm c}\approx 9.3$~km) 
and inner crust ($R_{\rm c}<r<R=10$~km) the rotating superfluid is organized
as an array of vortices parallel to the spin axis of the NS\footnote{The value $R_{\rm c}=9.3$~km is 
obtained assuming a $n=1$ polytropic density profile to describe the core and inner
crust, with $R = 10$~km and $M=1.4$~M$_\odot$ for the radius 
of the inner crust and mass of the neutron star (see \citealt{Pizzochero11}).}.
The vortices are pinned to the crustal lattice of ions, but are not
coupled with the normal component of the star.
Therefore, although the crust spins-down because of the electromagnetic braking,
the superfluid conserves its angular momentum\footnote{The angular velocity 
of the superfluid component $\Omega_{\rm s}$ is proportional to the number 
of vortices. As long as the vortices remain pinned to the lattice of ions, 
their number is conserved, hence $\Omega_{\rm s}$ also does not vary.}.
As the NS spins down, a rotational lag builds up between the superfluid vortices
and the normal component.
When it reaches a critical value, a hydrodynamical force acting on vortices,
the \emph{Magnus force}, unpins and moves them out.
The  \citet{Pizzochero11} model introduces for the first time a density profile of the pinning
force with maximum value $f_{\rm m}=1.1\times 10^{15}$~dyn~cm$^{-1}$ 
at densities $\rho_{\rm m}\approx 0.2\rho_{\rm 0}$, where $\rho_{\rm 0} = 2.8 \times 10^{14}$~g~cm$^{-3}$
is the nuclear saturation density.
Because of the shape of the density profile of the pinning force,
the critical rotational lag for depinning is maximum at $x_{\rm m} \approx 9.7$~km
(in cylindrical coordinates, with the $z$-axis parallel to the spin axis).
Therefore, vortices from $x<x_{\rm m}$ accumulates in a vortex sheet at $x_{\rm m}$.
When the rotational lag at $x_{\rm m}$ reaches the critical value, the vortex sheet suddenly
moves out and exchanges the stored angular momentum with the normal component, causing a glitch.

With this model it is possible to determine the fraction of vortices pinned
to the lattice nuclei in different regions of the interior of a neutron star
and the timescale of the pinning$-$unpinning process.
The model can be used to predict the interval
between glitches $\Delta t_{\rm gl}$, the jump in angular velocity $\Delta \Omega_{\rm gl}$
and the jump in angular acceleration $\Delta \dot{\Omega}_{\rm gl}$ during a glitch.

We leave the observables $\Omega$ and $\dot{\Omega}_\infty$ 
as well as the two free parameters of the model $f_{\rm m}$ and $Y_{\rm gl}$
free to vary. The parameter
$Y_{\rm gl}$ describes the fraction of vorticity coupled to the normal crust on timescales
of the glitch rise time ($\lesssim 40$~s; \citealt{Dodson02}).

According to \citet{Pizzochero11}, the interval between glitches
can be expressed as
\begin{equation} \label{eq. deltat_gl}
  \Delta t_{\rm gl} \simeq 28.8 \frac{R_6^2}{M_{1.4}}\frac{f_{15}}{|\dot{\Omega}_{-11}|} \mbox{ yr,}
\end{equation}
where $R_6=R_{\rm NS}/10^6$~cm, $M_{1.4}=M_{\rm NS}/1.4$M$_\odot$, 
$f_{15}=f_{\rm m}/10^{15}$~dyn~cm$^{-1}$, and $\dot{\Omega}_{-11}=\dot{\Omega}_\infty/10^{-11}$~rad~s$^{-2}$.

When $\dot{\Omega}_\infty$ increases, $\Delta t_{\rm gl}$ decreases
resulting in a larger probability that a glitch will be observed.
Indeed, the greater the spin-down rate of the crust, the more quickly
the critical rotational lag for depinning will be reached. 
In addition, Eq.  (\ref{eq. deltat_gl})  shows  that when $f_{\rm m}$ increases,
 $\Delta t_{\rm gl}$ also increases owing to the larger unpinning critical lag
required by the \emph{Magnus forces} acting on the neutron vortices 
to overcome the pinning force.

The jump in angular velocity
is given by \citep{Pizzochero11},
\begin{equation} \label{eq. delta omega_gl}
  \Delta \Omega_{\rm gl} \simeq 1.25 \times 10^{-4} \frac{Q_{0.95}R_6^2f_{15}}{M_{1.4}[1 - Q_{0.95}(1-Y_{0.05})]} \mbox{ rad s}^{-1} \mbox{ ,}
\end{equation}
where $Q_{0.95}=Q/0.95$ \citep{Zuo04}, $Q$ is the superfluid fraction of the star, 
and $Y_{0.05}=Y_{\rm gl}/0.05$ \citep{Pizzochero11}.

When $f_{\rm m}$ increases, $\Delta \Omega_{\rm gl}$ increases because of the 
larger number of neutron vortices unpinned simultaneously 
(see \citealt{Pizzochero11}). Equation (\ref{eq. delta omega_gl}) also shows that when 
$Y_{\rm gl}$ increases, $\Delta \Omega_{\rm gl}$ decreases because of the
reduced number of vortices storing angular momentum coupled 
to the normal component of the neutron star.

Since accreting pulsars usually have angular velocities
smaller than those of isolated pulsars, we expect  
the glitch rise time and the
recovery time after glitch\footnote{The post-glitch recovery stage, which follows the sudden spin-up,
is characterized by a quasi-exponential relaxation of the angular velocity
time derivative of the star toward the spin-down rate which the pulsar
would have had in the absence of the glitch
(\citealt{Baym69}; \citealt{Alpar84}; \citealt{Jones90}).}
to be longer in accreting pulsars.
Indeed, these timescales depend on the coupling timescale between the superfluid
and the normal component, which is inversely proportional to the
angular velocity of the NS:
\begin{equation} \label{eq timescales}
  \tau \propto \frac{1}{\Omega}
\end{equation}
(see, e.g., \citealt{Haskell14} and references therein).
Therefore, since the glitch rise time of Vela pulsar 
($\Omega_{\rm Vela}=70$~rad~s$^{-1}$) is $\tau_{\rm Vela}\approx1-10$~s \citep{Haskell12},
in a slow rotator such as GX~1+4 it would be $\tau_{\rm GX1+4}\approx 10^3-10^4$~s.
Similarly, the recovery timescales of slow rotators become longer as well.
The glitch is expected to appear as a simple step in angular velocity.

Glitches with relaxation timescales longer than those of Vela pulsar
have been observed in some slower anomalous X-ray pulsars (\citealt{Haskell14}; \citealt{Dib08}).
The glitch rise time has not yet been measured. An upper-limit of $\sim 40$~s 
has been placed by \citet{Dodson02} for Vela pulsar.

\subsection{Anti-glitch scenario}
\label{sect. anti-glitch}

Some XRBs show long-term spin-ups with timescales
on the order of $\sim10$ years (e.g., 12~yr of 4U~1626$-$67; 
\citealt{Chakrabarty97} and referencese therein),
and rates up to $\dot{\Omega} \approx 10^{-10}$~rad~s$^{-2}$
(e.g., KS~1947+300; \citealt{Galloway04}).
In principle they are good candidates  for  experiencing
the \emph{anti-glitch} scenario proposed by \citet{Pines80},
in which a pulsar can show a sudden spin-down
caused by a mechanism of angular momentum transfer similar to that of glitches.

\begin{figure}
\begin{center}
\includegraphics[bb=376 344 681 668,clip,width=\columnwidth]{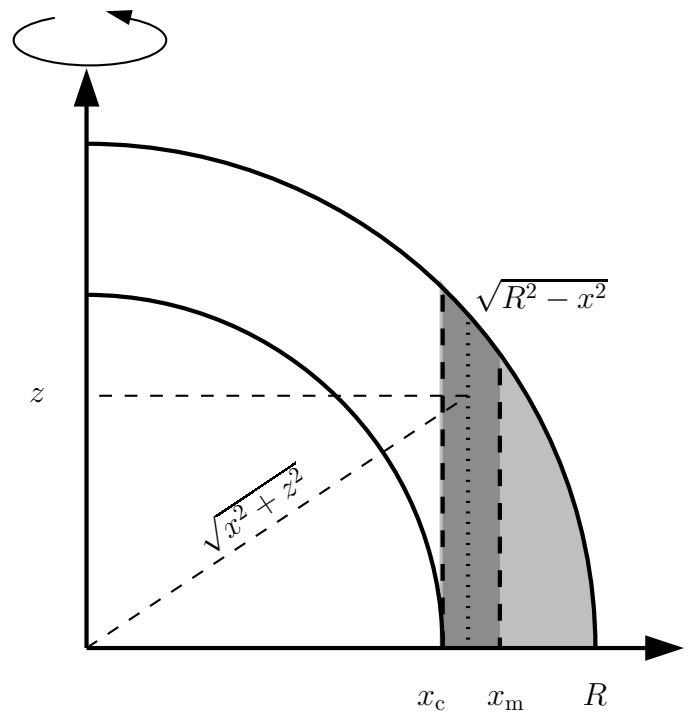}
\end{center}
\caption{Schematic representation of the cylindrical geometry assumed 
in Sect. \ref{sect. anti-glitch}.
The $z$-axis is parallel to the spin-axis of the NS. The dotted line shows
a vortex line with half length $\sqrt{R^2-x^2}$.}
\label{figure geometry}
\end{figure}

Here we adapt the snowplow model of \citet{Pizzochero11} 
to the scenario proposed by \citet{Pines80} to calculate the
expected jump in angular velocity caused by an anti-glitch $\Delta \Omega_{\rm a-gl}$.
Hereafter we will use cylindrical coordinates as shown in Fig. \ref{figure geometry}.

In the glitch scenario, when the pulsar slows down,
vortices from the core region are expelled outward, allowing the angular
momentum of the core to rebalance easily, while the vortices in excess
are accumulated at the boundary $x_{\rm c}$ and then expelled outward as
vortex sheet. In the anti-glitch scenario, on the contrary, the crust
accelerates. Before the anti-glitch, new vortices are created at $R$, near the equator,
as the star accelerates.
They are accumulated in a vortex sheet 
by the external depinning front\footnote{At any time $t$ before the glitch,
the external and internal depinning fronts represent the radial distance $x_{\rm ext}(t)$ 
and $x_{\rm int}(t)$  (with $x_{\rm ext}(t) > x_{\rm m}$ for the external depinning front and
$x_{\rm int}(t) < x_{\rm m}$ for the internal depinning front)
where the rotational lag $\Delta \Omega = \Omega_{\rm s}(x) - \Omega$ between
the local superfluid angular velocity $\Omega_{\rm s}(x)$ and that of the normal
component $\Omega$ is equal to the critical lag for depinning $\Delta \Omega_{\rm cr}$.
As the star spins up, a rotational lag builds up between the superfluid vortices
and the normal component. The critical lag for depinning is highest at $x_{\rm m}$.
Therefore, the external depinning front moves inward and the internal depinning
front outward, both to $x_{\rm m}$.
Since the NS spins up, the relative velocity $\vec{v}_{\rm L} - \vec{v}_{\rm s}$
of the vortex lines pinned to the lattice ($\vec{v}_{\rm L}$) with respect to
the superfluid ($\vec{v}_{\rm s}$) increases.
The Magnus force acting on the vortices is given by
\begin{equation} \nonumber
\vec{F}_{\rm m} = \kappa \rho_{\rm s} \vec{\hat{e}}_{\rm z} \times (\vec{v}_{\rm L} - \vec{v}_{\rm s}) \mbox{ ,}
\end{equation}
where $\vec{\hat{e}}_{\rm z}$ is the unit vector pointing along the rotation axis.
Because of the direction of $\vec{F}_{\rm m}$, the vortices unpinned by the
external and internal depinning fronts move inward
(while in the glitch scenario, they move outward).}
that moves inward from $R$ to $x_{\rm m}$.
In the meantime, as the internal depinning front
moves across the region $x<x_{\rm m}$, the vorticity moves
from the inner crust to the core and rebalances
the angular momentum in the core region.
Since the number of vortices within $x_{\rm m}$ does not change,
their accumulation in the core region must correspond to a depletion
in the regions of the inner crust with lower pinning potential,
which is around $x_{\rm c}$, the boundary between the two regions
(see \citealt{Pizzochero11} and figures 1 and 3 therein).
We assume that when the vortices accumulated at $x_{\rm m}$ 
by the external depinning front
are released simultaneously,
they fill the depleted region around $x_{\rm c}$. Hence,  
the transfer of angular momentum will take place in the region $x_{\rm c} \lesssim x \lesssim x_{\rm m}$.
We point out that the minimum ($x_{\rm min}$) and maximum ($x_{\rm max}$)
radii of the depletion region are unknown.
A precise estimate of these values requires a detailed study of vortex dynamics.
Nonetheless, it is still possible to roughly estimate 
the order of magnitude of the jump in angular
velocity caused by an anti-glitch relying on a few reasonable assumptions
on $x_{\rm min}$ and $x_{\rm max}$ as described below.

According to \citet{Seveso12}, the angular velocity of the superfluid component
of the star $\Omega_{\rm s}(x)$ is given by
\begin{equation} \label{omega_s}
\Omega_{\rm s}(x) = \frac{\kappa}{2\pi} \frac{N(x)}{x^2} \mbox{ ,}
\end{equation}
where $N(x)$ is the number of vortices in a cylindrical region of 
radius $x$, $\kappa = \pi \hbar/m_{\rm N}$, and $m_{\rm N}$ is the mass of a neutron.
Using Eq. (\ref{omega_s}), the number of vortices removed from the region
$x_{\rm m} < x < R$ and accumulated at $x_{\rm m}$ before the anti-glitch is
\begin{equation} \label{eq N_v}
  N_{\rm v} = \frac{2\pi}{\kappa} (R^2 - x_{\rm m}^2) \Delta \Omega_{\rm cr}(x_{\rm m}) \mbox{ ,}
\end{equation}
where $\Delta \Omega_{\rm cr}(x_{\rm m})$ is the critical rotational lag for depinning at $x_{\rm m}$.
Since we do not know  the exact size of the re-coupling region,
we calculate the angular momentum released during the anti-glitch $\Delta L_{\rm a-gl}$
based on two different assumptions.
In the first case we assume that $N_{\rm v}$ vortices
will be uniformly distributed in $x_{\rm min} < x < x_{\rm m}$
by integrating $N(x) = N_{\rm v} (x^2 - x_{\rm min}^2)/(x_{\rm m}^2 - x_{\rm min}^2)$,
where $x_{\rm min}$ corresponds to the minimum radius of the depletion zone 
around the boundary $x_{\rm c}$. 
From the definition $dL=\Omega_{\rm s}dI_{\rm s}$, $\Delta L_{\rm a-gl}$ can be expressed as
\begin{equation} \label{eq dl_agl}
  \Delta L_{\rm a-gl} = 2\kappa N_{\rm v} \int_{x_{\rm min}}^{x_{\rm m}} \frac{x^2 - x_{\rm min}^2}{x_{\rm m}^2 - x_{\rm min}^2}xdx \int_0^{\sqrt{R^2 - x^2}} \rho_{\rm s}(r)dz \mbox{ ,}
\end{equation}
where $r=\sqrt{x^2 + z^2}$ and
\begin{equation} \nonumber
\rho_s(r) = Q\frac{\pi M}{4R^3}\frac{sin(\pi r/R)}{\pi r/R}
\end{equation}
is the superfluid density profile \citep{Pizzochero11}.
In the second case, we assume that $N_{\rm v}$ vortices
will be uniformly distributed in $x_{\rm c} < x < x_{\rm max}$ ($x_{\rm max} < x_{\rm m}$),
where the pinning potential is lower.
In this case,  $\Delta L_{\rm a-gl}$ is
\begin{equation} \label{eq dl_agl2}
  \Delta L_{\rm a-gl} = 2\kappa \int_{x_{\rm c}}^{x_{\rm max}} N(x)xdx \int_0^{\sqrt{R^2 - x^2}} \rho_{\rm s}(r)dz \mbox{ ,}
\end{equation}
where
\begin{equation} \label{n_x}
N(x) = N_{\rm v}\times \left\{ \begin{array}{ll}
  (x^2 - x_{\rm c}^2)/(x_{\rm max}^2 - x_{\rm c}^2) & \quad \mbox{if $x_{\rm c} \leq x \leq x_{\rm max}$} \\
  1 & \quad \mbox{if $x_{\rm max}<x \leq x_{\rm m}$.}
       \end{array} \right.
\end{equation}
The jump in angular velocity of the normal component of the NS caused by the anti-glitch
is thus given by
\begin{equation} \label{eq domega_agl}
  \Delta \Omega_{\rm a-gl} = \frac{\Delta L_{\rm a-gl}}{I_{\rm tot}[1-Q(1-Y_{\rm gl})]} \mbox{ ,}
\end{equation}
where $I_{\rm tot}$ is the total momentum of inertia of the NS.
In Eqs. (\ref{eq dl_agl}) and (\ref{eq dl_agl2}) we integrate on the cylindrical regions
$x_{\rm min} < x < x_{\rm m}$ and $x_{\rm c} < x < x_{\rm max}$, 
where $x_{\rm min}$ and $x_{\rm max}$ are unknown.
Under the reasonable assumption that  $x_{\rm min}$ and $x_{\rm max}$
are close to $x_{\rm c}$, where the pinning potential of the inner crust is lower,
we find that $|\Delta \Omega_{\rm a-gl}|$ is approximately $10^{-5}-10^{-4}$~rad~s$^{-1}$
assuming $f_{\rm m}=1.1 \times 10^{15}$~dyn~cm$^{-1}$ and $Y_{\rm gl}=0.05$.

We note that Eq. (\ref{eq. deltat_gl}) can also be used to
describe the interval between anti-glitches.
Indeed, $\Delta t_{\rm gl}$ is the time required to reach the
critical rotational lag for depinning, no matter whether
the rotational lag between the superfluid vortices
and the normal component has been built through a long-term spin-down
or a long-term spin-up.
Hereafter we will use $\Delta t_{\rm gl}$ both for glitches and for anti-glitches.

\section{Observability}
\label{sect. observability}

%
The results of the previous section indicate that, in the case of accreting pulsars,
glitches (anti-glitches) should have 
maximum jumps in angular velocity $\Delta \Omega_{\rm gl} \approx 1.3 \times 10^{-4}$~rad~s$^{-1}$
($\Delta \Omega_{\rm a-gl} \approx 10^{-5}-10^{-4}$~rad~s$^{-1}$),
rise time $\tau_{\rm rise} \approx (10^2-10^3) \Omega^{-1}$~s, and long recovery timescales making them
appear as simple steps in angular velocity leaving $\dot{\Omega}$ nearly unchanged.

In accreting pulsars, accretion torque variations  (including sign reversals) result
in changes in the NS angular velocity. Several transitions between spin-up
and spin-down have been observed \citep{Nelson97}
and have been explained so far with sharp changes in the coupling between 
the matter flowing toward the NS and the magnetosphere.
Several models have been proposed both for disk-  and wind-fed pulsars
(e.g., \citealt{Zhang10} and references therein; \citealt{Wang81}).
However, a widely accepted model does not yet exist.
The timescale $t_\infty$ between two torque reversals   varies from source to source.
Some pulsars show numerous torque reversals with moderate spin period derivative
(e.g., Vela X$-$1, Her X$-$1; \citealt{Bildsten97}), others show quasi-periodic torque reversals
superimposed on a long-term spin-up trend (e.g., OAO~1657$-$415, \citealt{Barnstedt08};
EXO~2030+375, \emph{Fermi}/GBM\footnote{The \emph{Fermi} Team provides the pulse frequency history
of the accreting pulsars monitored with the Gamma-ray Burst Monitor (GBM) on board the \emph{Fermi}
satellite since its launch (June 11, 2008).
Preliminary results are available for download from the website 
\url{http://gammaray.nsstc.nasa.gov/gbm/science/pulsars.html}.}; 
SAX~J2103.5+4545, \citealt{Camero14}),
while others show steady spin-up or spin-down sporadically interrupted
by a torque reversal (e.g., GX~1+4, \citealt{Gonzalez-Galan12}; 
4U~1626$-$67, \citealt{Beri14}; 4U~1907+09, \citealt{Inam09}).

\begin{figure}
\begin{center}
\includegraphics[bb = 102 226 478 624,clip,width=\columnwidth]{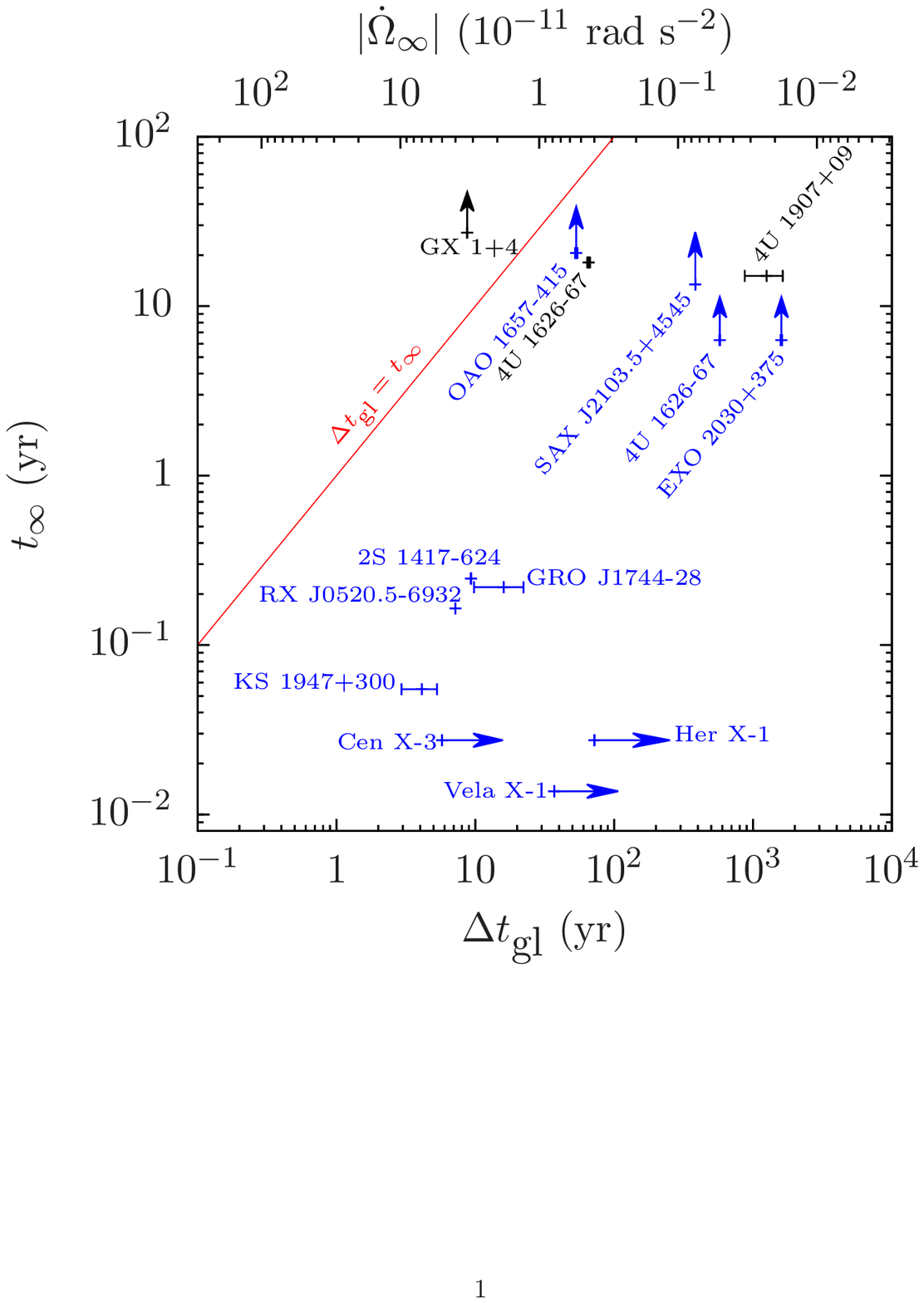}
\end{center}
\caption{$\Delta t_{\rm gl} - t_\infty$ diagram of 13 accreting pulsars.
Black  refers to spin-down, blue  to spin-up. Pulsars with horizontal arrows
experience a ``random walk'' of the spin period, and so  the value $\Delta t_{\rm gl}$ reported in the diagram is a lower-limit.
Pulsars with vertical arrows have ongoing spin-up or spin-down.
We obtained $t_\infty$ and $\Delta t_{\rm gl}(\dot{\Omega}_\infty)$ from previous works
(OAO~1657$-$415: \citealt{Jenke12}; 4U~1626$-$67: \citealt{Chakrabarty97}; \citealt{Beri14};
4U~1907+09: \citealt{Inam09}; SAX~J2103.5+4545: \citealt{Camero14}),
or from the \emph{Fermi}/GBM monitoring archive
(GX~1+4, EXO~2030+375, 2S~1417$-$624, RX~J0520.5$-$6932, 
GRO~J1744$-$28, KS~J1947+300, Cen~X$-$3, Vela~X$-$1, Her~X$-$1).}
\label{figure tinf}
\end{figure}

It is interesting to compare, for each pulsar, $t_\infty$ with $\Delta t_{\rm gl}$,
the time needed to reach the critical rotational lag for depinning.
The accreting pulsars with sufficiently long  time intervals of spin-down or spin-up
(i.e., with $t_\infty \gtrsim \Delta t_{\rm gl}$) are those more likely  to experience glitches or anti-glitches.
In Fig. \ref{figure tinf} we show $t_\infty$ and $\Delta t_{\rm gl}$ for a sample of accreting pulsars.
We calculate $\Delta t_{\rm gl}$ from $\dot{\Omega}_\infty$ using Eq. (\ref{eq. deltat_gl}) and
obtain $t_\infty$ and $\dot{\Omega}_\infty$ from previous works or from the \emph{Fermi}/GBM
monitoring archive. 
Black symbols refer to the values observed during a long-term
spin-down, while those obtained during a long-term spin-up are plotted in blue  
(for some pulsars, e.g., 4U~1626$-$67,  both spin-up and spin-down episodes have been observed, hence both cases
are indicated).
For those pulsars that exhibit quasi-periodic torque reversals
superimposed on a long-term spin-up, namely OAO~1657$-$415, EXO~2030+375, and SAX~J2103.5+4545,
we use $t_\infty$ and $\Delta t_{\rm gl}$ related to the long-term spin-up.
For persistent pulsars that show a ``random walk'' of the spin period
with several torque reversals and moderate $|\dot{\Omega}_\infty|$
(Vela~X$-$1, Cen~X$-$3, Her~X$-$1), we give the lower-limits on $\Delta t_{\rm gl}$.
We show the lower-limits on $t_\infty$ for pulsars with spin-up
or spin-down that are still ongoing (GX~1+4, OAO~1657$-$415, SAX~J2103.5+4545,
spin-up of 4U~1626$-$67, and EXO~2030+375).
GX~1+4 has $t_\infty > \Delta t_{\rm gl}$, making it the best candidate for observing
glitches. Other good candidates are 4U~1626$-$67 and OAO~1657$-$415, with $t_\infty \approx \Delta t_{\rm gl}$.
We point out that,  in principle, all the accreting pulsars
can experience glitches or anti-glitches, although with a lower probability. Indeed, they have long spin period evolutions ($10^4-10^6$~yr)
during which they might have built up a rotational lag for depinning
close to the critical value.

While glitches are traditionally observed with  radio observations, 
the numerous examples of glitches  seen in magnetars clearly show that they can also be detected  
by X-ray observatories  (e.g., \citealt{Kaspi00}). 
The size of the jumps in angular velocity
observed in the magnetars is the same or smaller than that expected in accreting pulsars
($\Delta \Omega \approx 10^{-7}-10^{-4}$~rad~s$^{-1}$, \citealt{Sasmaz13} and references therein). 

A suitable spacing of the observations, allowing  phase-connected timing 
ephemeris to be derived, is required in order  to detect glitches and to distinguish them 
from other timing irregularities induced by variations in the accretion torque. 
The observation of correlated changes in the source flux should help 
to recognize accretion-induced torque variations.
Existing  instruments with a wide field of view, such as the \emph{Fermi}/GBM,  
\emph{Swift}/BAT, and \emph{INTEGRAL}/IBIS are already providing
repeated observations of  the brightest X-ray pulsars that could be used to search for glitches, 
but the observing strategy of these satellites has not been optimized in this respect. 
In the future, a relevant contribution in this field could be given by the 
\emph{LOFT} mission, proposed in the context of the 
European Space Agency science program \citep{Feroci14}. 
The Large Area Detector (LAD) instrument on \emph{LOFT} will be able to detect the pulsations  
of bright sources like 4U~1626$-$67 or GX~1+4 in observations of only a few kiloseconds. 
A program of suitably spaced monitoring observations will be easily implemented, 
also exploiting the frequent coverage of the Galactic plane with the LOFT Wide Field X-ray monitor.

\section{Conclusions}

We use starquake and superfluid vortex models
to outline for the first time the expected observational 
properties of glitches in accreting pulsars.

We find that glitches caused by quakes of the crust in an accreting slow pulsar
are very rare and their detection unlikely.
On the contrary, glitches caused by the transfer of angular momentum between
the superfluid and the normal components may take place more often.
In the superfluid vortex scenario they can have
maximum jump in angular velocity
$\Delta \Omega_{\rm gl} \approx 1.3 \times 10^{-4}$~rad~s$^{-1}$.
Some accreting pulsars show long-term spin-up periods
during which they may undergo anti-glitches.
Therefore, we modify the snowplow model of \citet{Pizzochero11} on the basis
of the anti-glitch scenario proposed by \citet{Pines80} 
to determine the expected properties of anti-glitches in these objects.
We find that anti-glitches in accreting pulsars have maximum jumps
in angular velocity 
of about the same order of magnitude or ten times smaller than $\Delta \Omega_{\rm gl}$.
Both glitches and anti-glitches are expected to have long rise times
($\tau_{\rm rise} \approx 10^2-10^3\Omega^{-1}$~s) and
long recovery timescales, with the glitch appearing 
as a simple step in angular velocity.

By comparing the timescales between two consecutive torque reversals 
$t_\infty$ and the interglitch timescale $\Delta t_{\rm gl}$ of a sample of 
accreting pulsars, we find  that GX~1+4 has $t_\infty > \Delta t_{\rm gl}$ (see Fig. \ref{figure tinf}).
It is therefore the best candidate for observing glitches.
Other good candidates with $t_\infty \approx \Delta t_{\rm gl}$
are 4U~1626$-$67 and OAO~1657$-$415.
These sources can be monitored to search for glitches
and anti-glitches with the 
currently operating X-ray instruments and represent good targets
for future missions devoted to X-ray timing such as the proposed \emph{LOFT}.

\begin{acknowledgements}
We thank the anonymous referee for constructive comments that
helped to improve the paper.
This work is partially supported by the Bundesministerium f\"ur
Wirtschaft und Technologie through the Deutsches Zentrum f\"ur Luft
und Raumfahrt (grant FKZ 50 OG 1301).
VD and AS thank the Deutsches Zentrum f\"ur Luft- und Raumfahrt (DLR) and
Deutsche Forschungsgemeinschaft (DFG) for financial support
(grant DLR 50 OR 0702).
Partial support comes from NewCompStar, COST Action MP1304.
Part of this work is based on the publicly available ``GBM Accreting Pulsar Histories''
provided by the \emph{Fermi} team.
\end{acknowledgements}

\bibliographystyle{aa} 
\bibliography{lducci_spinevol}

\end{document}